\documentclass[dvips]{acta}
\usepackage{supertabular,lscape,epsfig}
\usepackage{amssymb}
\usepackage{amsmath}
\usepackage{float}
\mathchardef\mhyphen="2D
\usepackage{placeins}%\Floatbarier
\DeclareSymbolFont{ppa}{OT1}{ppl}{m}{it}
\DeclareMathSymbol{\vv}{\mathalpha}{ppa}{'166}

\SetPages{1}{17}

\SetVol{69}{2019}

\begin{document}
\newcommand\pvalue{\mathop{p\mhyphen {\rm value}}}
%Zwarte naglowki, jeden wiersz, appendix
\newcommand{\TabApp}[2]{\begin{center}\parbox[t]{#1}{\centerline{
  {\bf Appendix}}
  \vskip2mm
  \centerline{\small {\spaceskip 2pt plus 1pt minus 1pt T a b l e}
  \refstepcounter{table}\thetable}
  \vskip2mm
  \centerline{\footnotesize #2}}
  \vskip3mm
\end{center}}

%Zwarte naglowki, jeden wiersz
\newcommand{\TabCapp}[2]{\begin{center}\parbox[t]{#1}{\centerline{
  \small {\spaceskip 2pt plus 1pt minus 1pt T a b l e}
  \refstepcounter{table}\thetable}
  \vskip2mm
  \centerline{\footnotesize #2}}
  \vskip3mm
\end{center}}

%Zwarte naglowki, dwa wiersze
\newcommand{\TTabCap}[3]{\begin{center}\parbox[t]{#1}{\centerline{
  \small {\spaceskip 2pt plus 1pt minus 1pt T a b l e}
  \refstepcounter{table}\thetable}
  \vskip2mm
  \centerline{\footnotesize #2}
  \centerline{\footnotesize #3}}
  \vskip1mm
\end{center}}

%Zwarte naglowki, jeden wiersz, appendix
\newcommand{\MakeTableApp}[4]{\begin{table}[p]\TabApp{#2}{#3}
  \begin{center} \TableFont \begin{tabular}{#1} #4
  \end{tabular}\end{center}\end{table}}

%Zwarte naglowki, jeden wiersz
\newcommand{\MakeTableSepp}[4]{\begin{table}[p]\TabCapp{#2}{#3}
  \begin{center} \TableFont \begin{tabular}{#1} #4
  \end{tabular}\end{center}\end{table}}

%Zwarte naglowki, jeden wiersz
\newcommand{\MakeTableee}[4]{\begin{table}[htb]\TabCapp{#2}{#3}
  \begin{center} \TableFont \begin{tabular}{#1} #4
  \end{tabular}\end{center}\end{table}}

%Zwarte naglowki, dwa wiersze
\newcommand{\MakeTablee}[5]{\begin{table}[htb]\TTabCap{#2}{#3}{#4}
  \begin{center} \TableFont \begin{tabular}{#1} #5
  \end{tabular}\end{center}\end{table}}

%Tabela w okreslonym miejscu
\newcommand{\MakeTableH}[4]{\begin{table}[H]\TabCap{#2}{#3}
  \begin{center} \TableFont \begin{tabular}{#1} #4
  \end{tabular}\end{center}\end{table}}

%Tabela w okreslonym miejscu, zwatre nag³ówki, jeden wiersz
\newcommand{\MakeTableHH}[4]{\begin{table}[H]\TabCapp{#2}{#3}
  \begin{center} \TableFont \begin{tabular}{#1} #4
  \end{tabular}\end{center}\end{table}}

\newfont{\bb}{ptmbi8t at 12pt}
\newfont{\bbb}{cmbxti10}
\newfont{\bbbb}{cmbxti10 at 9pt}
\newcommand{\uprule}{\rule{0pt}{2.5ex}}
\newcommand{\douprule}{\rule[-2ex]{0pt}{4.5ex}}
\newcommand{\dorule}{\rule[-2ex]{0pt}{2ex}}
\def\thefootnote{\fnsymbol{footnote}}
\begin{Titlepage}

\Title{Over 78\,000 RR~Lyrae Stars in the Galactic Bulge and Disk\\
from the OGLE Survey\footnote{Based on observations obtained with 
the 1.3-m Warsaw telescope at the Las Campanas Observatory of the 
Carnegie Institution for Science.}}
\Author{I.~~S~o~s~z~y~ñ~s~k~i$^1$,~~
A.~~U~d~a~l~s~k~i$^1$,~~
M.~~W~r~o~n~a$^1$,~~
M.\,K.~~S~z~y~m~a~ñ~s~k~i$^1$,\\
P.~~P~i~e~t~r~u~k~o~w~i~c~z$^1$,~~
J.~~S~k~o~w~r~o~n$^1$,~~
D.~~S~k~o~w~r~o~n$^1$,~~
R.~~P~o~l~e~s~k~i$^{2,1}$,\\
S.~~K~o~z~³~o~w~s~k~i$^1$,~~
P.~~M~r~ó~z$^{1,3}$,~~
K.~~U~l~a~c~z~y~k$^{4,1}$,~~
K.~~R~y~b~i~c~k~i$^1$,~~
P.~~I~w~a~n~e~k$^1$\\
and~~M.~~G~r~o~m~a~d~z~k~i$^1$
}
{$^1$Astronomical Observatory, University of Warsaw, Al.~Ujazdowskie~4, 00-478~Warszawa, Poland\\
e-mail: soszynsk@astrouw.edu.pl\\
$^2$Department of Astronomy, Ohio State University, 140 W. 18th Ave., Columbus, OH~43210, USA\\
$^3$Division of Physics, Mathematics, and Astronomy, California Institute of Technology, Pasadena, CA 91125, USA\\
$^4$Department of Physics, University of Warwick, Gibbet Hill Road, Coventry, CV4~7AL,~UK}
\Received{December 9, 2019}
\end{Titlepage}

\Abstract{We present an upgrade of the OGLE Collection of RR~Lyrae stars in
the Galactic bulge and disk. The size of our sample has been doubled and
reached 78\,350 RR~Lyr variables, of which 56\,508 are fundamental-mode
pulsators (RRab stars), 21\,321 pulsate solely in the first-overtone (RRc
stars), 458 are classical double-mode pulsators (RRd stars), and 63 are
anomalous RRd variables (including five triple-mode pulsators). For all the
newly identified RR~Lyr stars, we publish time-series photometry obtained
during the OGLE Galaxy Variability Survey. 

We present the spatial distribution of RR~Lyr stars on the sky, provide a
list of globular clusters hosting RR~Lyr variables, and discuss the
Petersen diagram for multimode pulsators. We find new RRd stars belonging
to a compact group in the Petersen diagram (with period ratios $P_{\rm
  1O}/P_{\rm F}\approx0.74$ and fundamental-mode periods $P_{\rm
  F}\approx0.44$~d) and we show that their spatial distribution is roughly
spherically symmetrical around the Milky Way center.}{Stars: variables:
RR~Lyrae -- Stars: oscillations -- Galaxy: bulge -- Galaxy: disk --
Catalogs}

\Section{Introduction}
RR~Lyrae stars play a crucial role in understanding the formation,
composition, and kinematics of the Galactic halo and thick disk. These
radially pulsating horizontal-branch stars, with periods ranging from 0.2
to 1~d, belong to the very old stellar population (older than
$\approx10$~Gyr). RR~Lyr variables constitute the most numerous group of
classical pulsating stars, follow a relatively narrow period--luminosity
relation and have distinctive light curves, which make them primary
distance indicators among old, low-mass stars.

In recent years, various wide-field sky surveys -- LINEAR (Sesar \etal
2013), Catalina Sky Survey (Drake \etal 2014), Pan-STARRS (Hernitschek
\etal 2016, Sesar \etal 2017), SuperWASP (Greer \etal 2017), ASAS-SN
(Jayasinghe \etal 2018, 2019, 2020), Gaia (Holl \etal 2018, Clementini
\etal 2019) -- have published extensive catalogs of RR~Lyr stars in the
Milky Way. These samples have been used to investigate the structural
properties of the Galactic halo, including the exploration of stellar
streams composed of material that has been tidally striped from dwarf
galaxies (\eg Torrealba \etal 2015, 2019, Hernitschek \etal 2018, Koposov
\etal 2019). However, due to crowding and blending, the Galactic bulge and
disk region was either intentionally avoided by the variability surveys or
the published catalogs of variable stars were affected by significantly
reduced levels of completeness and purity.

The Optical Gravitational Lensing Experiment (OGLE) is a photometric sky
survey devoted to long-term monitoring of densely populated regions, such
as the Galactic bulge and disk. So far, OGLE published a catalog of more
than 39\,000 RR~Lyr stars identified over 182 square degrees toward the
central parts of the Milky Way (Soszyñski \etal 2011, 2014,
2017). Additionally, 45 RR~Lyr stars were found by Pietrukowicz \etal
(2013) in the OGLE pilot survey of the Galactic disk. The OGLE Collection
of Variable Stars (OCVS) has enabled us to gain deep insight into the
structure of the Milky Way bulge (Pietrukowicz \etal 2015), but also to
explore the structure of the Sagittarius Dwarf Galaxy (SgrDG, Hamanowicz
\etal 2016, Ferguson and Strigari 2019). Accurate and well-sampled OGLE
light curves have been used to investigate a variety of phenomena
occurring in RR~Lyr stars: additional radial and non-radial modes, the
Blazhko effect, switching of the pulsation modes, binarity, Oosterhoff
dichotomy (\eg Netzel \etal 2015, 2018, Smolec \etal 2016, Prudil \etal
2017, 2019ab, Das \etal 2018, Netzel and Smolec 2019). Moreover, the highly
complete OGLE samples of variable stars have been commonly used for the
validation of the catalogs obtained from other projects (\eg Braga \etal
2019, Clementini \etal 2019, Plachy \etal 2019).

In 2013, the OGLE team launched a new observational subproject named the
Galaxy Variability Survey (GVS), with the main goal to photometrically
monitor stars in the Galactic disk and outer bulge. The GVS is shallower
than the regular OGLE survey (exposure time of 25~s \vs 100--150~s in
the regular OGLE project), but it covers a much larger area of the sky --
about 3000~square degrees in total. The multi-epoch photometric data
collected by the GVS project have so far been used by Udalski \etal (2018)
to discover over a thousand new classical, type~II, and anomalous Cepheids
in the Milky Way. The extended catalog of Cepheids was used by Mróz
\etal (2019) to derive an accurate rotation curve of our Galaxy and by
Skowron \etal (2019) to construct a three-dimensional map of the Galactic
disk and to study the recent star formation history in the Milky Way.

In this paper, we present the collection of RR~Lyr stars identified among
over a billion stars regularly observed by the GVS project in the Galactic
bulge and disk. Our new sample doubles the number of Galactic RR~Lyr stars
included so far in the OCVS. The completeness and purity of our collection
is very high, which makes it complementary to other large RR~Lyr catalogs
produced by wide-field surveys.

The paper is organized as follows. An overview of the OGLE photometric data
is given in Section~2. In the following section, we briefly review the
selection and classification criteria of the RR~Lyr stars. The structure of
the collection of RR~Lyr stars is described in Section~4. Section~5
presents the analysis of the completeness of the resulting sample and its
comparison with other large-scale catalogs of RR~Lyr stars. Globular
cluster members and multi-mode RR~Lyr variables are discussed in the
following two sections. Finally, Section~8 summarizes our results.

\Section{Observations and Data Reduction}
The OGLE project is a large-scale, long-term, two-band ({\it I}- and {\it
V}-band) photometric survey aimed at systematic exploration of the
variable sky. Since 2010 (when the fourth phase of the OGLE survey,
OGLE-IV, started), the project has used 32-chip mosaic CCD camera, with a
field of view of 1.4 square degrees and a pixel scale of 0\zdot\arcs26,
mounted on the 1.3-m Warsaw Telescope at Las Campanas Observatory in Chile
(the Observatory is operated by the Carnegie Institution for Science).

The OGLE GVS subsurvey has been carried out since 2013. Together with the
regular OGLE-IV survey, the GVS project monitors brightness of about two
billion stars located in the area of more than 3000~square degrees along
the Galactic plane. The GVS footprint spans about 230~degrees in the
Galactic longitude and reaches 7~degrees above and below the Galactic plane
in the disk region and $\pm15$~degrees in the bulge region. At the end of
2019, images covering 2750~square degrees in the Galactic disk and bulge
(including the regular OGLE fields) have already been reduced and this work
presents the collection of RR~Lyr stars detected in this region. The
remaining GVS fields will be reduced soon, and variable stars identified in
this region will be added to the OCVS.

The majority of the GVS observations were obtained through the Cousins {\it
I}-band filter with the exposure time of 25~s. The {\it V}-band light
curves, currently unavailable for most of the stars, will be added to the
collection in the future. The time series collected by GVS project are not
as well sampled as those from the regular OGLE observations. The total
number of points per light curve ranges from about 20 to 200, with a median
value of 112~epochs. The timespan ranges from several months to several
years, with a typical value of 3~yr.

The GVS project is about 1~mag shallower than the regular OGLE survey, so
the saturation limit of the GVS photometry is at about 11~mag, while the
faintest stars for which useful information about their variability can be
obtained have $I\approx19.5$~mag. For further details on the GVS project,
data reduction, and astrometric calibrations we refer the reader to the
papers by Udalski \etal (2015, 2018).

\Section{Selection and Classification of RR~Lyr Stars}
Our search for RR~Lyr stars in the GVS fields was carried out in two
stages. An initial stage of the selection was described by Udalski \etal
(2018). Briefly, stars that left significant residua on the difference
images produced by the OGLE reduction pipeline were selected. A visual
inspection of their light curves in conjunction with the Fourier analysis
and positions in the color--magnitude diagram allowed us to detect more
than one thousand previously unknown Cepheids and several thousand RR~Lyr
stars (Udalski 2017).

We began the second stage of our variable stars' selection from the massive
period search for all GVS {\it I}-band light curves with at least 15 data
points. For this purpose, we applied the {\sc Fnpeaks}
code\footnote{http://helas.astro.uni.wroc.pl/deliverables.php?lang=en\&active=fnpeaks}
which uses the Discrete Fourier Transform algorithm. The {\sc Fnpeaks} code
provides the most significant periods with amplitudes and signal-to-noise
ratios. We have performed a period analysis for more than a billion light
curves.

Then, objects with periods shorter than 50~d and the largest
signal-to-noise ratio of their periodicities were initially classified by
using light-curve template fitting. We followed the analysis undertaken by
Soszyñski \etal (2019) for variable stars in the Magellanic Clouds. The
algorithm is based on the carefully selected light curves of pulsating
stars and eclipsing binary systems previously detected by the OGLE team in
the Galactic bulge and Magellanic Clouds. We chose the best sampled
single-periodic light curves representing a wide range of variability
types. Each light curve was folded with the pulsation or orbital period and
binned into 1000~bins, while its amplitude was rescaled to 1~mag. Example
template light curves were showed in the papers by Soszyñski \etal (2016a,
2019).

For each star analyzed by the template fitting program, we obtained the
most probable variability type with a parameter reflecting the goodness of
the fit. Then, the best candidates for pulsating stars were revised during
the visual inspection of their light curves. Besides RR~Lyr stars, we
detected, among others, new Cepheids, $\delta$~Sct stars, long-period
variables, eclipsing binaries, and various types of rotating variables. Our
classification was based largely on the light-curve characteristics,
including the parameters of the {\it I}-band light curve Fourier
decomposition. Candidates for double-mode RR~Lyr stars were preselected on
the basis on their period ratios and their light curves were also subject
to visual inspection. The final phase of our selection procedure was a
match of the OGLE database with other large catalogs of RR~Lyr stars in the
Milky Way (Section~5), which additionally brought approximately
600~RR~Lyr variables overlooked in the previous stages of our variability
search.

As a result, we identified 39\,232 {\it bona fide} RR~Lyr stars. We divided
this sample into four classes: fundamental-mode RRab stars, first-overtone
RRc stars, double-mode RRd stars, and anomalous multi-mode (double- and
triple-mode) RRd stars. This classification was based on the pulsation
periods, light curve morphology, and period ratios (in the case of
multi-mode pulsators).

Some RRc stars can have almost symmetric light curves with morphologies
similar to those of close binary systems phased with half their orbital
period. To minimize the contamination from close binary systems, we
required perceptible skewness of a light curve to classify a star as an RRc
variable. This requirement may reduce the completeness of our sample of RRc
stars, especially for fainter objects (Section~5).

\vspace{5mm}
\Section{The OGLE Collection of Galactic RR~Lyr Stars}
RR~Lyr variables identified in the GVS fields have been added to the
 OCVS. Now, the OGLE sample contains in total 78\,350 Galactic RR~Lyr
 stars: 56\,508 RRab, 21\,321 RRc, and 521 RRd stars (including 63
 anomalous RRd stars). The complete list of the OGLE Galactic RR~Lyr stars
 along with their {\it I}-band and {\it V}-band light curves (if available)
 are provided online through a user-friendly WWW interface or FTP sites:
\begin{center}
{\it http://ogle.astrouw.edu.pl}\\
{\it ftp://ftp.astrouw.edu.pl/ogle/ogle4/OCVS/blg/rrlyr/}\\
{\it ftp://ftp.astrouw.edu.pl/ogle/ogle4/OCVS/gd/rrlyr/}\\
\end{center}

The newly identified variables have extended the lists of previously
published RR~Lyr stars in the Galactic bulge (Soszyñski \etal 2014, 2017)
and Galactic disk (Pietrukowicz \etal 2013). We have kept the naming
convention used in the OCVS: OGLE-BLG-RRLYR-NNNNN and OGLE-GD-RRLYR-NNNNN
(for bulge and disk RR~Lyr stars, respectively), where NNNNN is a
five-digit consecutive number. For all stars, we provide their basic
properties: pulsation modes, J2000 equatorial coordinates, periods,
intensity mean magnitudes, amplitudes, and parameters of light curve
Fourier decomposition. The pulsation periods were improved with the {\sc
Tatry} code by Schwarzenberg-Czerny (1996). For each star we provide a
$60\arcs\times60\arcs$ finding chart.

\begin{landscape}
\begin{figure}[p]
  \centerline{\includegraphics[bb = 100 20 480 800, clip, width=11cm, angle=-90]{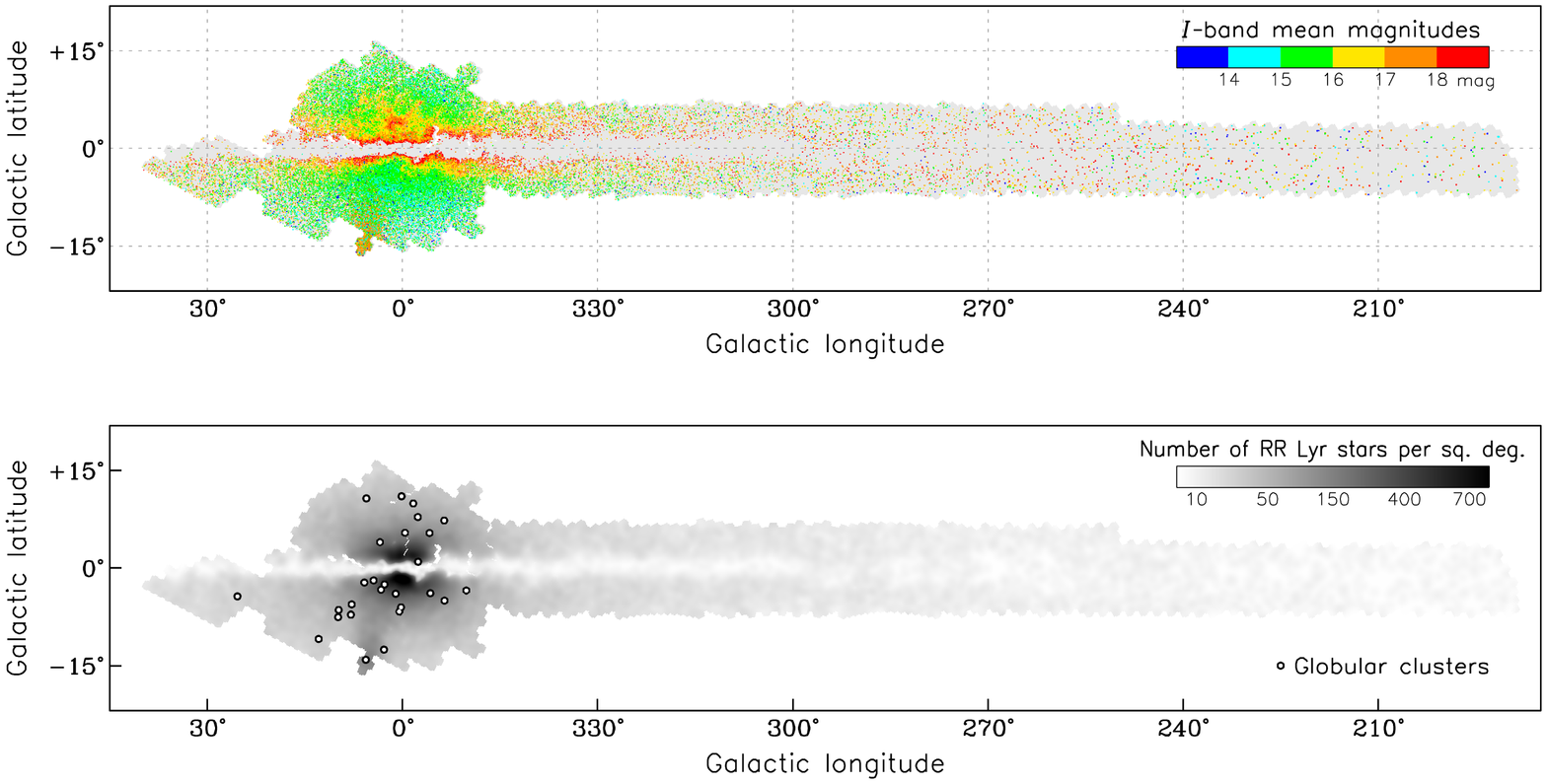}} 
\vskip9pt 
\FigCap{Distribution of Galactic RR~Lyr stars in Galactic coordinates. In
  the {\it upper panel} the colors of the points indicate magnitudes of
  each star, following the color code in the panel. The gray area shows the
  OGLE-IV footprint. Lower panel presents a surface density map of RR~Lyr
  stars. We used the square-root scaling to show the full dynamic range of
  the map. White circles indicate the positions of 27 globular clusters
  hosting RR~Lyr stars.}
\end{figure}
\end{landscape}

We also upgraded light curves and parameters of 762 Galactic RR~Lyr stars
already included in the OCVS, but not observed so far during the regular
OGLE-IV project. Most of these variables were discovered from the OGLE-III
observations (Soszyñski \etal 2011, Pietrukowicz \etal 2013). The OGLE
collection of RR~Lyr stars will continue to grow as we discover new
variables in the newly reduced fields monitored by the GVS survey.

\Subsection{Spatial Distribution}
Fig.~1 presents the distribution of RR~Lyr variables on the sky. In the
upper panel, different colors of the points represent different {\it
  I}-band mean magnitudes of the stars. As one can see, in the Galactic
bulge and in the adjacent section of the disk, RR~Lyr stars become
gradually fainter in the lines-of-sight closer to the Galactic plane. This
is a result of heavy interstellar extinction in this area. For this reason,
most of RR~Lyr stars located in the regions closest to the Galactic equator
are not accessible to the OGLE survey. These highly obscured areas have
been explored in the near-infrared domain by the VISTA Variables in V{\'i}a
L{\'a}ctea (VVV) survey (Gran \etal 2016, Minniti \etal 2017, Contreras
Ramos \etal 2018, Majaess \etal 2018).

In the lower panel of Fig.~1, we present a surface density map of our
RR~Lyr sample. We used a square root scaling to show the full dynamic range
this map. The surface density ranges from less than one RR~Lyr star per
square degree in the regions close to the Galactic anticenter to about 1000
RR~Lyr variables in the most populated fields in the bulge. In the lower
panel of Fig.~1, we also present positions of 27 globular clusters hosting
RR~Lyr stars (Section~6).

\begin{figure}[t]
\includegraphics[bb = 30 440 580 760, clip, width=13.1cm]{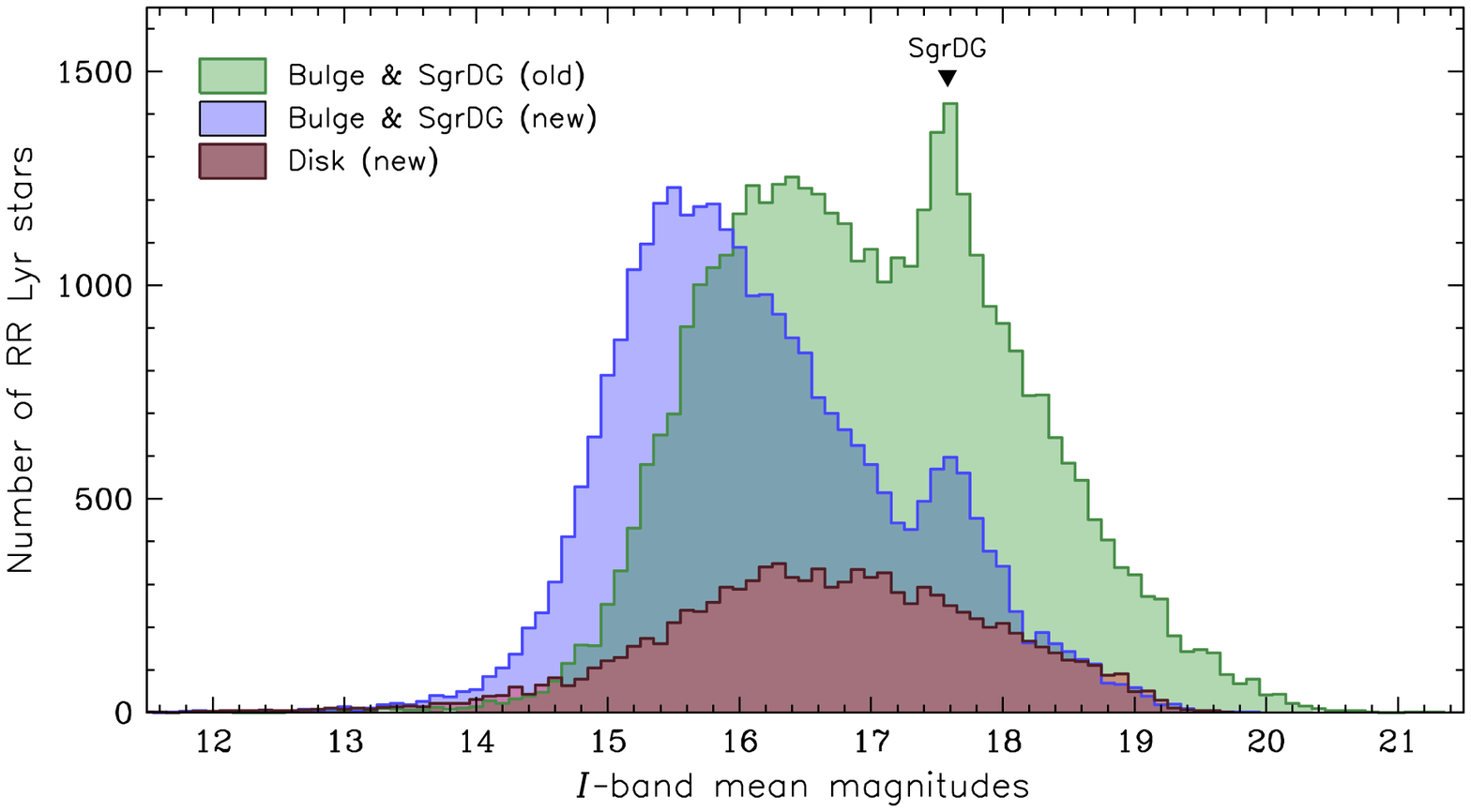}
\FigCap{Distributions of the mean {\it I}-band magnitudes of RR~Lyr stars
in the Galactic bulge and disk. Different colors of the histograms show
three samples: green -- RR~Lyr stars found by Soszyñski \etal (2011, 2014,
2017) toward the Galactic bulge in the regular OGLE fields, violet --
objects detected toward the bulge in the in the GVS fields, brown -- RR~Lyr
variables identified in the GVS fields in the Galactic disk.}
\end{figure}

\Subsection{Magnitude Distribution}
Our collection contains a significant number of RR~Lyr stars belonging to
the SgrDG. They are well visible in Fig.~2, where the distributions of the
apparent mean {\it I}-band magnitudes of the Galactic RR~Lyr stars are
shown. The three histograms display: magnitude distribution of the RR~Lyr
stars found by Soszyñski \etal (2011, 2014, 2017) in the regular OGLE
fields toward the Galactic bulge (``old'' variables), RR~Lyr variables
detected in the outer bulge observed by the GVS (``new''), and RR~Lyr stars
identified in the Galactic disk fields. The SgrDG members produce a peak
around a magnitude of 17.6~mag in the first two of these distributions.

The ``new'' variables in the central regions of the Milky Way are on
average brighter than the ``old'' ones, which is a result of a much smaller
interstellar extinction in the outer bulge than in the inner bulge,
monitored by the regular OGLE survey. Different depths of the GVS and
regular OGLE project can be seen at the edges of the distributions -- the
faintest RR~Lyr stars detected from the regular survey have mean {\it
I}-band luminosities of about 20.5~mag -- approximately one magnitude
fainter than the GVS detection limit. On the other hand, the saturation
limit is respectively brighter in the GVS dataset.

\Section{Completeness and Cross-Check with Other Catalogs} 
The completeness and purity levels of our collection of RR~Lyr stars are
affected by various factors. First of all, we draw the reader's attention
to the fact that the OGLE-IV mosaic camera has technical gaps between CCD
chips, which decrease the absolute completeness of our sample by
6--7\%. Objects that fell into these ``dead zones'' of our detector have not
been observed at all by the OGLE survey.

Considering only those objects that have been observed by OGLE, the
completeness of our collection is a strong function of the temporal
sampling, brightness, amplitudes, presence of the Blazhko effect, and, most
of all, the pulsation modes of RR~Lyr stars. The completeness and purity
levels of our collection are definitely largest for RRab stars, since they
have characteristic light curves that are not easily mistaken for other
types of periodic variables. We also expect that the efficiency of our
search for RRd variables is larger than for RRc stars, because two periods
with a certain ratio unambiguously identify an object as a double-mode
pulsator.

In order to assess the completeness of our sample, we used stars located in
the overlapping parts of adjacent GVS fields. Such objects have double
entries in the OGLE database, so we had the opportunity to detect them
twice during our selection process (however, the final version of our
collection contains only one entry per star -- usually the one with larger
number of data points). Because stars located near the edges of the fields
are often affected by a smaller number of epochs, we took into
consideration only those light curves that consisted of more than 30 data
points. We {\it a posteriori} checked that 1735 RR~Lyr stars in our
collection have their duplicates in other GVS fields, so we had a chance to
find 3470 counterparts. We independently detected 3289 of them, which gives
the completeness level of more than 94\% for the whole sample. For RRab
variables only, the completeness is about 96\%, while for RRc stars it is
equal to 91\%.

Using the same method, we estimated the efficiency of our RR~Lyr search in
various luminosity ranges. The completeness level for RRc variables drops
below 80\% for objects fainter than $I=17$~mag and below 30\% for stars
fainter than $I=18$~mag. For RRab stars, the completeness level is still
above 80\% for objects with the mean luminosity around 18~mag.

To improve the completeness of the RR~Lyr collection, we cross-matched our
initially identified sample with a number of extensive catalogs containing
Galactic RR~Lyr variables: the International Variable Star Index (VSX,
Watson \etal 2006), ASAS (\eg Pojmañski 2002), ASAS-SN (Jayasinghe \etal
2018, 2019, 2020), SuperWASP (Greer \etal 2017), Catalina Sky Survey (Drake
\etal 2014), ATLAS (Heinze \etal 2018), VVV (Gran \etal 2016, Minniti \etal
2017, Contreras Ramos \etal 2018, Majaess \etal 2018), Pan-STARRS
(Hernitschek \etal 2016, Sesar \etal 2017), and Gaia Data Release 2 (Holl
\etal 2018, Clementini \etal 2019). For some of these surveys (ASAS,
Catalina Sky Survey, SuperWASP), we found only several common RR~Lyr stars,
due to small overlaps between sky footprints or observed magnitude
ranges. By combining our sample with other lists of RR~Lyr stars (VSX,
ASAS-SN, ATLAS, VVV, Pan-STARRS), we found from about 2000 to 9000 matches,
and the largest overlap -- over 15\,000 RR~Lyr stars -- was found with the
Gaia DR2 catalog. In total, 23\,604 of RR~Lyr stars from our collection
($\approx60\%$) have been found in at least one of the explored catalogs,
so approximately 40\% of our sample are likely new detections.

We also carefully studied objects overlooked during our selection and
classification process (Section~3), but classified as RR~Lyr variables in
the above catalogs. We found nearly 40\,000 such objects in the OGLE GVS
database (most of them from the Pan-STARRS catalog, but also over 5000
RR~Lyr candidates from the Gaia DR2 sample) and we visually inspected their
OGLE light curves. This exercise yielded extra 604 RR~Lyr stars, which were
added to the final version of our collection increasing its
completeness. The majority of these variables were missing from the
preliminary version of the OCVS due to their poorly-sampled or noisy light
curves. We also overlooked some RRc stars that suffer various selection
effects. We must also notice that a number of RR~Lyr candidates from
external catalogs cannot be unambiguously categorized using OGLE light
curves, because of sparse photometry. On the other hand, among RR~Lyr
candidates we also found thousands of stars obviously belonging to other
variability types: eclipsing binaries, Cepheids, $\delta$~Sct stars,
spotted variables, or sometimes constant stars.

\Section{RR~Lyr Stars in Globular Clusters}
It has long been known that RR~Lyr variables provide firm constraints to
study properties of globular clusters and other ancient stellar systems. In
the paper describing the OGLE collection of RR~Lyr stars in the Galactic
bulge (Soszyñski \etal 2014), we provided a list of 15~globular clusters
hosting RR~Lyr variables. In this work, we expand this list by adding 12
other globular clusters located in the OGLE GVS fields, including M62
(NGC~6266) -- one of the most RR~Lyr-rich globular cluster in the Galaxy
(Contreras \etal 2010). The sky positions of the all globular clusters
containing RR~Lyr stars are plotted in the lower panel of Fig.~1.

\MakeTableee{
l@{\hspace{8pt}} c@{\hspace{6pt}} c@{\hspace{6pt}} c@{\hspace{8pt}}
c@{\hspace{6pt}} c@{\hspace{6pt}}}{12.5cm}{Globular clusters containing RR~Lyr stars}{\hline \noalign{\vskip3pt}
\multicolumn{1}{c}{Cluster name} & RA & Dec & Cluster & $N_{\rm RR}$ & $N_{\rm fieldRR}$ \\
  & (J2000) & (J2000) & radius [\arcm] & &
 (estimated) \\
\noalign{\vskip3pt}
\hline
\noalign{\vskip3pt}
 M62 (NGC~6266) & 17\uph01\upm13\ups & $-30\arcd06\arcm44\arcs$ & 15.0 & 210 &  2.8 \\
 NGC~6284       & 17\uph04\upm29\ups & $-24\arcd45\arcm53\arcs$ &  6.2 &   8 &  0.6 \\
 NGC~6287       & 17\uph05\upm09\ups & $-22\arcd42\arcm29\arcs$ &  4.8 &   4 &  0.3 \\
 NGC~6293       & 17\uph10\upm10\ups & $-26\arcd34\arcm54\arcs$ &  8.2 &   7 &  1.9 \\
 M9 (NGC~6333)  & 17\uph19\upm12\ups & $-18\arcd30\arcm59\arcs$ & 12.0 &  13 &  1.6 \\
 NGC~6380       & 17\uph34\upm28\ups & $-39\arcd04\arcm09\arcs$ &  3.6 &   9 &  0.3 \\
 M28 (NGC~6626) & 18\uph24\upm33\ups & $-24\arcd52\arcm12\arcs$ & 11.2 &  21 &  2.3 \\
 NGC~6638       & 18\uph30\upm56\ups & $-25\arcd29\arcm47\arcs$ &  7.3 &  22 &  1.0 \\
 NGC~6642       & 18\uph31\upm54\ups & $-23\arcd28\arcm35\arcs$ &  5.8 &  23 &  0.6 \\
 M70 (NGC~6681) & 18\uph43\upm13\ups & $-32\arcd17\arcm31\arcs$ &  8.0 &   5 &  0.4 \\
 NGC~6712       & 18\uph53\upm04\ups & $-08\arcd42\arcm22\arcs$ &  9.8 &  13 &  0.6 \\
 NGC~6717       & 18\uph55\upm06\ups & $-22\arcd42\arcm03\arcs$ &  5.4 &   3 &  0.1 \\
\noalign{\vskip3pt}
\hline}

Table~1 summarizes the basic parameters of the newly added clusters: their
coordinates, radii \footnote{The parameters of the globular clusters have
been taken from the web page:\\ {\it
http://messier.obspm.fr/xtra/supp/mw\_gc.html}}, and numbers of RR~Lyr
stars detected within one cluster angular radius from the cluster
center. In the last column of Table~1, we provide expected numbers of field
RR~Lyr stars lying by chance in the same line-of-sight as globular
clusters. These numbers were estimated by counting RR~Lyr variables in the
rings around each cluster. In Table~1, we included only those clusters that
contain at least three more RR~Lyr stars than the expected number of field
variables inside an area outlined by the cluster radius. In the FTP sites of
the collection, we provide files {\sf gc.dat} containing 668 RR~Lyr stars
found in the area occupied by globular clusters. Note that some of these
variables may not be physically related to the clusters, but the
discrimination between field and cluster RR~Lyr stars is beyond the scope
of this paper.

\begin{figure}[t]
\includegraphics[bb = 30 240 580 760, clip, width=13.1cm]{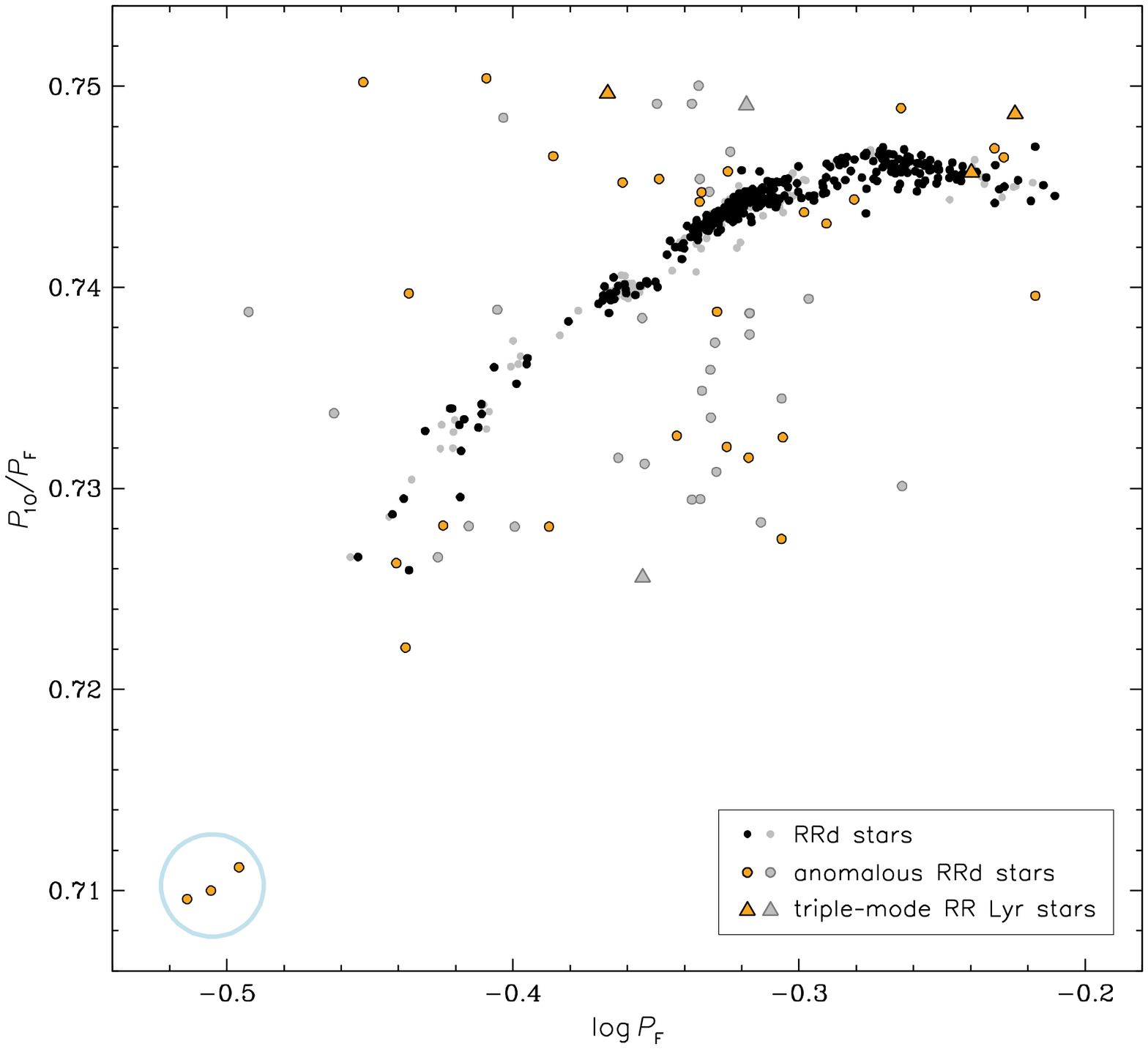}
\FigCap{Petersen diagram for Galactic RRd stars in the OCVS. Black and
  yellow points indicate objects included in this upgrade of the
  collection, while gray points mark multimode RR~Lyr stars published by
  Soszyñski \etal (2011, 2014, 2017). Large light-blue circle marks the
  position of three short-period double-mode anomalous RRd stars (their
  light curves are shown in Fig.~4).}
\end{figure}

\Section{Multimode RR~Lyr Stars}
RR~Lyr variables oscillating simultaneously in two or three radial modes
(RRd stars) are a rich source of information on stellar pulsation and
evolution theories. One of the OGLE results in this field was the discovery
of a new subclass of multi-mode RR~Lyr variables, named anomalous RRd stars
(Soszyñski \etal 2016b). With this upgrade of the OCVS, the number of known
anomalous RRd stars in the Milky Way has increased by a factor of two (from
31 to 63). In turn, the sample of Galactic ``classical'' RRd stars in the
OGLE collection has been tripled: from 148 to 458 objects.

This disproportionate increase in the number of RRd variables in relation
to the total sample of RR~Lyr stars in our collection, is due to the fact
that the incident rate of double-mode RR~Lyr stars seems to be a strong
function of metallicity. RRd stars constitute only about 0.5\% of all
RR~Lyr variables in the Galactic bulge, but the relative number of RRd
stars is gradually increasing with the distance from the Milky Way center
and exceeds 3\% around Galactic anti-center. For comparison, the incident
rates of RRd stars in the Large and Small Magellanic Clouds are about 5\%
and 10\%, respectively.

Fig.~3 shows the Petersen diagram, in which the period ratio, $P_{\rm
1O}/P_{\rm F}$, is plotted against the logarithm of the fundamental-mode
period $P_{\rm F}$. The Petersen diagram is a sensitive probe of masses and
chemical content of double-mode RR~Lyr stars, because positions of
double-mode pulsators in this diagram strongly depend on stellar properties
(\eg Popielski \etal 2000). Gray symbols in Fig.~3 indicate multimode
RR~Lyr stars in the Galactic bulge published by Soszyñski \etal (2011,
2014, 2017), while black and yellow symbols are new detections. Our
classification of the anomalous RRd stars, was largely based on their
position in the Petersen diagram, but we also took into account ratios of
amplitudes associated to both pulsation modes, light curve morphology, and
presence of the Blazhko effect (Soszyñski \etal 2016b).

We also found three new triple-mode variables (marked with triangles in
Fig.~3). Following Soszynski \etal (2017), we classified these triple-mode
pulsators as anomalous RRd stars. Table~2 gives the complete list of five
triple-mode RR Lyr variables included in the OCVS.

\renewcommand{\arraystretch}{1.05}
\MakeTableee{lcccccc}{12.5cm}{Triple-Mode RR Lyr stars in the galactic bulge and disk}
{\hline
\noalign{\vskip3pt}
\multicolumn{1}{c}{Identifier} & $P_{\rm F}$ & $A_{\rm F}$ & $P_{\rm 1O}$ & $A_{\rm 1O}$ & $P_{\rm 2O|3O}$ & $A_{\rm 2O|3O}$ \\
 &  [days] & [mag] & [days] & [mag] & [days] & [mag] \\
\noalign{\vskip3pt}
\hline
\noalign{\vskip3pt}
OGLE-BLG-RRLYR-24137 & 0.4419265 & 0.154 & 0.3206446 & 0.125 & 0.2199141 & 0.056 \\
OGLE-BLG-RRLYR-38791 & 0.4804440 & 0.194 & 0.3598870 & 0.117 & 0.2865899 & 0.043 \\
OGLE-BLG-RRLYR-57262 & 0.5964392 & 0.199 & 0.4465113 & 0.161 & 0.3558438 & 0.045 \\
OGLE-GD-RRLYR-05935  & 0.4296175 & 0.108 & 0.3220625 & 0.090 & 0.2561699 & 0.039 \\
OGLE-GD-RRLYR-09200  & 0.5758730 & 0.215 & 0.4294370 & 0.127 & 0.3415855 & 0.058 \\
\hline}

\begin{figure}[t]
\includegraphics[bb = 60 330 600 760, clip, width=14cm]{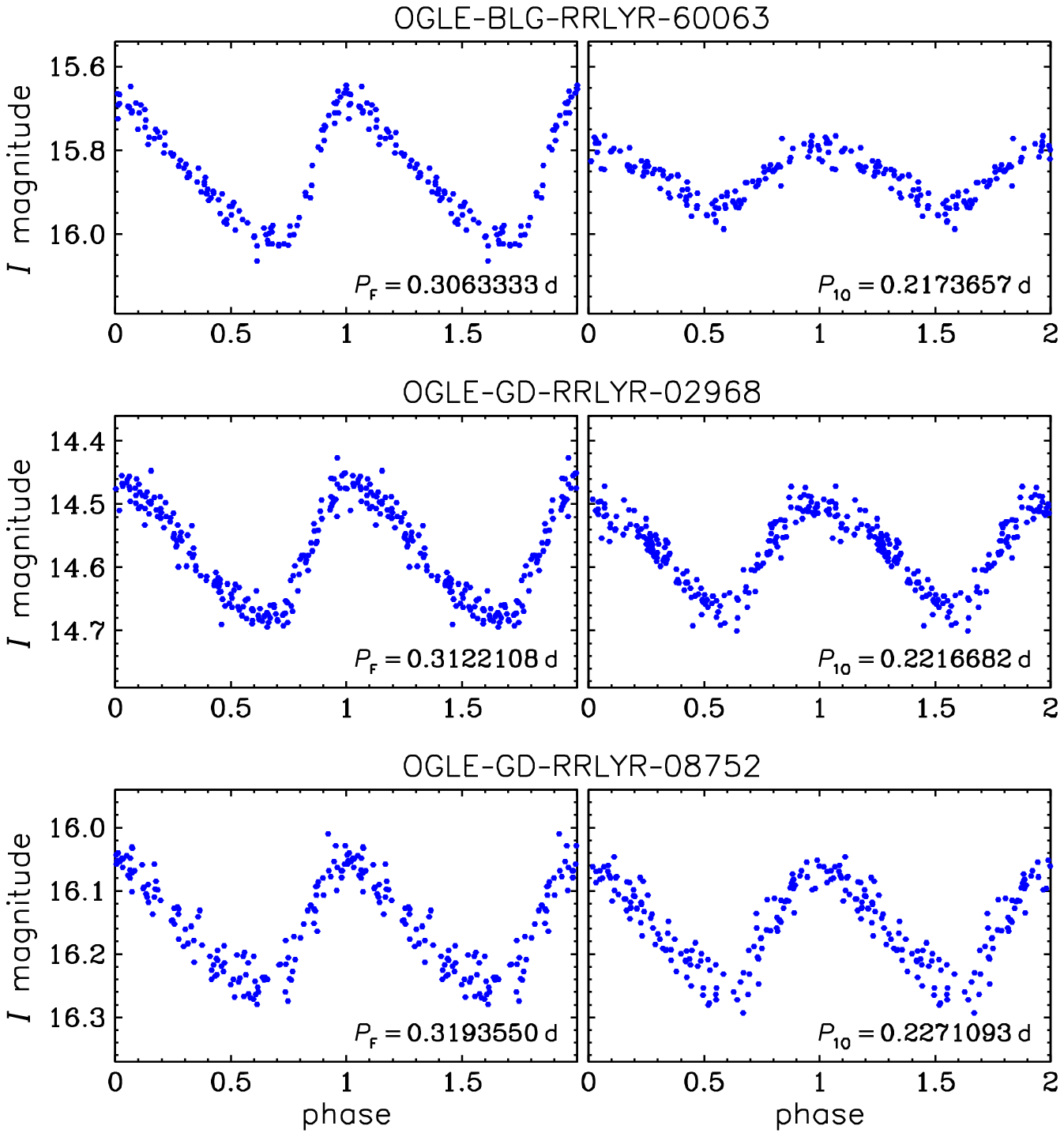}
\FigCap{Disentangled {\it I}-band light curves of three short-period
double-mode RR Lyr stars marked with a blue circle in Fig.~3.}
\end{figure}

At the bottom of the Petersen diagram, the blue circle marks three
double-mode pulsators with exceptionally low period ratios of around
0.71. We also classified these stars as anomalous RRd variables. Their
disentangled light curves are shown in Fig.~4. Note that both periods have
relatively large amplitudes in each of the three objects, which indicates
that both periods are related to the radial (fundamental-mode and
first-overtone) pulsation modes.

It is worth noting that Fig.~3 does not show all double-periodic variables
in our collection. At least six variables have secondary periods shorter by
a factor of 0.68--0.71 than the primary period. In most such cases the
amplitudes of the additional modes are significantly smaller than the
amplitudes of the dominant modes. Double-mode RR~Lyr stars of the same type
were studied by Prudil \etal (2017) based on the OCVS. They considered
several possibilities (radial modes in unusually high metallicity RR~Lyr
stars, binary evolution pulsators, non-radial modes) and found no
convincing scenarios explaining the nature of the secondary periodicities
in these objects. Since Prudil \etal (2017) suggested that the primary
periods in these objects correspond to the fundamental-mode pulsations, we
tentatively classify them as RRab variables, while their secondary periods
are given in the remarks in our catalog.

\Subsection{Group of RRd Stars with $P_{\rm 1O}/P_{\rm F}\approx0.74$}
During the preparation of the OGLE-III catalog of RR~Lyr stars in the
Galactic bulge (Soszyñski \etal 2011), we paid attention to a distinct
group of 16 double-mode variables that are clumped around $P_{\rm
  F}\approx0.44$~d and $P_{\rm 1O}/P_{\rm F}\approx0.74$ in the Petersen
diagram. ~Soszyñski \etal (2014) extended this list to 28 objects ~and~ no-

\begin{landscape}
\begin{figure}[p]
\centerline{\includegraphics[bb = 80 20 560 760, clip, width=12cm, angle=-90]{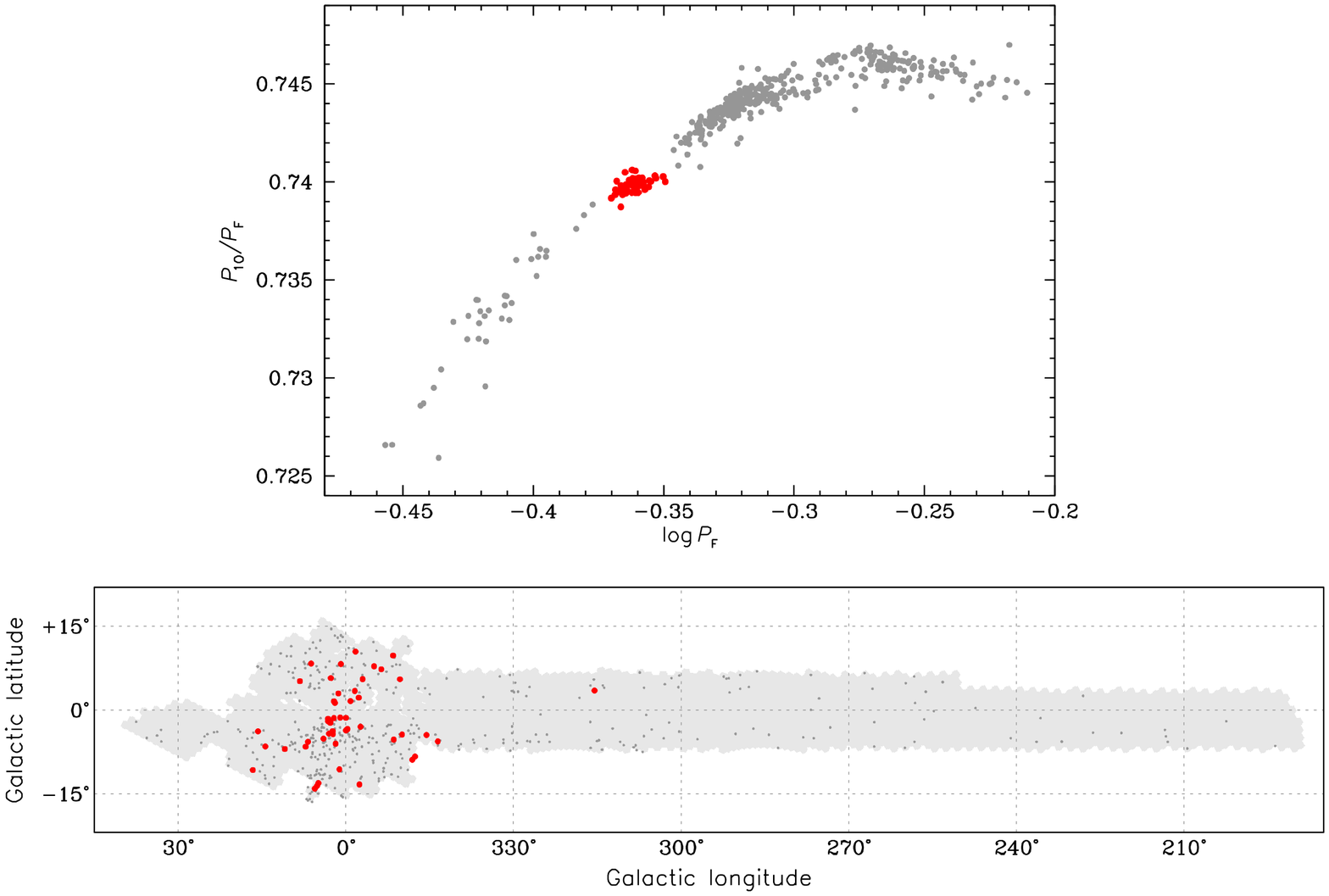}} 
\FigCap{{\it Upper panel:} Petersen diagram for ``classical'' RRd
  stars. Red points show the compact group of 50 RRd stars with $P_{\rm
  F}\approx0.44$~d and $P_{\rm 1O}/P_{\rm F}\approx0.74$. {\it Lower panel:}
  spatial distribution of the same stars on the sky. The gray area shows
  the OGLE-IV footprint.}
\end{figure}
\end{landscape}

\noindent ticed that their spatial distribution is consistent with a
stellar stream that nearly vertically crosses the bulge. It suggested that
these stars belong to a relic of a disrupted cluster or dwarf
galaxy. However, Kunder \etal (2019) found that these pulsationally clumped
RRd stars have a large range of radial velocities and proper motions, which
indicates that they do not belong to a coherently moving group of stars as
would be expected for stellar streams.

In the upgraded collection, we increase the number of double-mode RR~Lyr
variable with $P_{\rm F}\approx0.44$~d and $P_{\rm 1O}/P_{\rm
F}\approx0.74$ by additional 22 stars. In the Petersen diagram (red
circles in the upper panel of Fig.~5), these objects are still clumped in a
separate compact group. The sky distribution of all these stars is shown in
the lower panel of Fig.~5. All but one of these stars are located in the
Galactic bulge and are roughly symmetrically distributed around the
Galactic center. Thus, our extended sample of RRd stars confirms the
conclusion of Kunder \etal (2019) that the stellar stream turned out to be
an illusion resulted from a small number statistics and incomplete coverage
of the Galactic bulge by the regular OGLE-IV fields.

Nonetheless, these 50 RRd stars with period ratios around 0.74 are clumped
in a separate group in the Petersen diagram, which may be a signature of
their common origin. It is likely that these stars are indeed relics of a
cluster or galaxy accreted by the Milky Way, but any kinematic signature
has now been lost. Such a scenario would be in agreement with the
hierarchical merging model for Milky Way formation.

\Section{Conclusions}
We carried out a systematic search for RR~Lyr stars in the OGLE GVS fields
covering Galactic disk and outer bulge. We identified more than 39\,000
RR~Lyr stars, which doubles the number of Galactic RR~Lyr stars included in
the OCVS. The explored area of the sky was increased by a factor of 15
compared to the regular OGLE footprint in the Galactic bulge.

About 40\% of the newly identified RR~Lyr stars have not been included in
any existing catalog of pulsating stars, but the greatest advantage of our
collection is an exceptionally high level of completeness (94\%) and a very
low contamination. Our sample is complementary to other large catalogs of
RR~Lyr stars in the Milky Way halo. The astrophysical potential of the OGLE
collection of RR~Lyr stars in the Galactic bulge and disk cannot be
overstated. It allows one to study three-dimensional structure of the old
stellar component in the vicinity of the Galactic plane and in the
bulge-halo transition region, to answer the question what fraction of the
sample is inherent to the bulge and disk and what fraction is just passing
by from the halo population, to explore the abundance patterns in the Milky
Way, to investigate globular clusters, and interstellar matter
distribution. We are also convinced that OGLE time-series photometry will
be useful in a variety of asteroseismological analyses.

\Acknow{We would like to thank M. Kubiak, G. Pietrzyñski, £.~Wy\-rzykowski
and M. Pawlak for their contribution to the collection of the OGLE
photometric data presented in this paper. We thank Z.~Ko³aczkowski and
A.~Schwar\-zenberg-Czer\-ny for providing software used in this study.

This work has been supported by the National Science Centre, Poland,
grant MA\-ESTRO no. 2016/22/A/ST9/00009. The OGLE project has received
funding from the Polish National Science Centre grant MAESTRO 
2014/14/A/ST9/00121. This research has made use of the International
Variable Star Index (VSX) database, operated at AAVSO, Cambridge,
Massachusetts, USA.}

\end{document}